\documentclass[preprint,showpacs,preprintnumbers,amsmath,amssymb,nofootinbib]{revtex4}
\usepackage{graphicx}
\usepackage{dcolumn} 
\usepackage{bm}      

\newcommand{\bea}{\begin{eqnarray}}
\newcommand{\eea}{\end{eqnarray}}

\begin{document}
\title{Statefinder  parameters for quintom  dark energy model}         
\author{  Puxun Wu and
Hongwei Yu \footnote{Corresponding author} }

\address
{ Department of Physics and Institute of  Physics,\\ Hunan Normal
University, Changsha, Hunan 410081, China}


\begin{abstract}
We perform   in this paper a  statefinder diagnostic to a dark
energy model with  two scalar fields, called ``quintom", where one
of the scalar fields has a canonical kinetic energy term and the
other has a negative one. Several kinds of potentials  are
discussed. Our results show that the statefinder diagnostic can
differentiate quintom model with other dark energy models.
\end{abstract}

\pacs{98.80.Es,98.80.Cq, 98.80.Jk}

 \maketitle

It is widely believed nowadays that the present universe is
undergoing an accelerating expansion. The converging evidences
come from the analysis of data from supernova~\cite{Per,Per1},
CMB~\cite{Spe,Spe1,Spe2,Spe3,Spe4,Spe5} and WMAP~\cite{Bah,Bah1}.
In order to explain the cosmic accelerating expansion,  a modified
theory of gravity or the existence of  dark energy is required.
Perhaps the simplest
 candidate of dark energy is the cosmological constant with equation
of state $w=p/\rho=-1$. However, there exist two problems with it,
i.e., why the cosmological constant is so tiny compared to the
theoretical expectations and why the remnant cosmological constant
is becoming visible precisely at the present time in the cosmic
history and why it does not exactly vanish. The inspiration coming
from inflation has suggested that dark energy models are likely to
be described by the dynamics of a (multi) scalar field(s), such as
quintessence~\cite{Peeb,Peeb1,Peeb2,Peeb3,Peeb4,Peeb5,Peeb6,Peeb7,Peeb8,Peeb9,Peeb10,
Peeb11,Peeb12,Peeb13,Peeb14,Peeb15,Peeb16} and
phantom~\cite{Cal,Cal1,Cal2,Cal3,Cal4,Cal5,Cal6,Cal7,Cal8,Cal9,Cal10,Cal11,Cal12,
Cal13,Cal14,Cal15,Cal16,Cal17,Cal18,Cal19,Cal20, Cal201,
Cal21,Cal22,Cal23,Cal24,Cal25,Cal26,Cal27,Cal28,Cal29,Cal30,Cal31,
Cal32,Cal33,Cal34,Cal35,Cal36}. The quintessence scalar field has a
positive kinetic term in its Lagrangian, which violates the strong
energy condition but not the dominant energy condition, and the
evolution of its equation of state parameter $w$  is in the range of
$-1 \leq w \leq 1$.  The phantom scalar field on the other hand
possesses  a negative kinetic term in its Lagrangian, which leads to
some strange properties, such as the violation of the dominant
energy condition and  the occurrence of Big
rip~\cite{Cald,Cald1,Cald2,Cald3,Cald4,Cald5,Cald6,Cald7,Cald8,Cald9}(
Let us note here that the possible avoidance of the Big Rip by
various kinds of physical effects has been discussed by many
authors~\cite{wu,wu1,wu2,wu3,wu4,wu5,wu6,wu7}.). The evolution of
the equation of state parameter  $w$ for phantom is in the range of
$ w < -1$. Other models including the Chaplygin
gas~\cite{Kamen,Kamen1}, braneworld
models~\cite{Dvali,Dvali1,Dvali2,Dvali3,Dvali4}, holographic
models~\cite{LiM} {\it et al} are also proposed to account for the
present accelerating cosmic expansion.

The recent analysis of various cosmological data seems to indicate
that it is mildly favored that the evolution of the equation of
state parameter of the dark energy changes from $w > -1$ to $w
<-1$ at small redshift\cite{UAlam,UAlam1,UAlam2,UAlam3}. Although
this result may be considered as tentative only since the standard
$\Lambda$ CDM model remains inside $\approx 2\sigma$ error bars
for all data, it is worthwhile to explore theoretical models which
can bring $w$ crossing -1.  In this regard, most models mentioned
above can not do the job. ``Quintom"~\cite{Fengb,Fengb1,Guo},
which assumes that the dark energy is composed of quintessence and
phantom, on the other hand, can implement $w$ crossing -1 and in
some cases fits the cosmological data better than  models with
$w\geq -1$. In addition "quintom" model can reconcile the
coincidence problem and predicts some interesting features in the
evolution and fate of our universe~\cite{Fengb,Fengb1}.
Considering a flat homogeneous and isotropic FRW universe filled
with matter, radiation and ``quintom" dark energy which is
composed of a phantom scalar field $\phi$ and a normal scalar
field $\varphi$, the dynamic equations for the universe, phantom
and normal scalar fields can be described by
 \bea
 \frac{\ddot{a}}{a}=\frac{8\pi G}{3}\bigg[\dot{\phi}^2-\dot{\varphi}^2+V(\phi,\varphi)
 -\frac{\Omega_{mat0}}{2a^3} -\frac{\Omega_{rad0}}{2a^4}\bigg]\;,
 \eea
 \bea
 \ddot{\phi}+3\frac{\dot{a}}{a}\dot{\phi}-V(\phi,\varphi),_{\varphi}=0\;,
 \eea
 \bea
 \ddot{\varphi}+3\frac{\dot{a}}{a}\dot{\varphi}+V(\phi,\varphi),_{\varphi}=0\;,
 \eea
 where
$V(\phi,\varphi)$ is the interacting potential between phantom and
quintessence, $0$ denotes   the present time, a dot means a
derivative with respect to time $t$ and $a$ is the scale factor.
The energy density and pressure for the quintom can be expressed
as
  \bea
 \rho_{qui}= -\dot{\phi}^2+\dot{\varphi}^2+V(\phi,\varphi)\;,
  \eea
  \bea
 p_{qui}= -\dot{\phi}^2+\dot{\varphi}^2-V(\phi,\varphi)\;.
  \eea
Thus we can obtain the equation of state for the quintom
 \bea
w_{qui}=\frac{p_{qui}}{\rho_{qui}}=\frac{-\dot{\phi}^2+\dot{\varphi}^2-V(\phi,\varphi)}
{-\dot{\phi}^2+\dot{\varphi}^2+V(\phi,\varphi)}\;.
 \eea
 This expression  implies $w_{qui}\geq-1$ when $|\dot{\varphi}|
\geq |\dot{\phi}|$ and $w_{qui}<-1$ when $|\dot{\varphi}|
<|\dot{\phi}|$.

Since  there are more and more models proposed  to explain the
cosmic acceleration, it is very desirable to find a way to
discriminate between the various contenders in a model independent
manner. In this regard,  Sahni {\it et al.}~\cite{Sahni1,Sahni}
recently proposed a cosmological diagnostic pair $\{r, s\}$ called
statefinder,  which are defined as
 \bea\label{state}
 r\equiv \frac{\dddot{a}}{aH^3}\;,\qquad
 s\equiv\frac{r-1}{3(q-1/2)}\;,
 \eea
  to differentiate between different forms of dark energy.
Here $q$ is the deceleration parameter.  Apparently  statefinder
parameters only depend on  $a$,  and is thus a geometrical
diagnostic.  It is easy to see that statefinder
 differentiates the expansion dynamics with  higher derivatives of
 scale factor $a$ and is a natural next step beyond $H$ and  $q$.
  Since different cosmological models
involving dark energy exhibit qualitatively different evolution
trajectories in the $s-r$ plane, this statefinder diagnostic can
 differentiate
 various kinds of dark energy models. For LCDM (or $\Lambda$CDM) cosmological
model, which
 consists  of a mixture of vacuum
 energy and cold  dark mass, the
 statefinder parameters correspond to a fixed point  $\{r=1, s=0\}$.
  By far some models,  including the cosmological
constant, quintessence, phantom, the Chaplygin gas, braneworld
models, holographic models,  interacting and coupling dark energy
models ~\cite{Sahni,Alam, Gori, Zimd, Zhang, Zhangx,WuYu}, have
been successfully differentiated.  For example, although the
quintessence model with inverse power law potential, the phantom
model with power law potential and the Chaplygin gas models all
tend to approach the LCDM fixed point,  for quintessence and
phantom models the trajectories lie in the regions $s > 0$, $r <
1$ while for Chaplygin  gas models the trajectories lie in the
regions $s < 0$, $r> 1$, and on the other hand the quintessence
tracker models and the Chaplygin gas models have typical
trajectories similar to arcs of a parabola (upward and downward)
respectively while for phantom model the trajectories are
different~\cite{Alam, Gori, Sahni, WuYu}. For the coupled
quintessence models the trajectories of $r(s)$
 form swirl before reaching the attractor~\cite{Zhang}.

In this  paper we apply the statefinder  diagnostic to the quintom
dark energy model. To begin with,  let us use another form of
statefinder parameters which can be written as
 \bea\label{statefinder}
 r=1+ \frac{9}{2}\frac{(\rho+p)}{\rho}\frac{\dot{p}}{\dot{\rho}}\;,\qquad
 s=\frac{(\rho+p)}{\rho}\frac{\dot{p}}{\dot{\rho}}\;.
\eea
 Here $\rho$  is the total energy density  and $p$ is the total
pressure in the universe.  Since the total energy, quintom energy
and radiation
 are conserved respectively, we have $\dot{\rho}=-3H(\rho+p)$,
$\dot{\rho}_{qui}=-3H(1+w_{qui})\rho_{qui}$ and
$\dot{\rho}_{rad}=-4H\rho_{rad}$. Thus we can obtain
 \bea\label{rs}
 r=1- \frac{3}{2}[\dot{w}_{qui}/H-3w_{qui}(1+w_{qui})]\Omega_{qui}+2\Omega_{rad}\;,
 \eea
 \bea
 s=\frac{-3[\dot{w}_{qui}/H-3w_{qui}(1+w_{qui})]\Omega_{qui}+4\Omega_{rad}}{9w_{qui}\Omega_{qui}+3\Omega_{rad}}\;,
\eea
  and
\bea q=\frac{1}{2}(1+3w_{qui}\Omega_{qui}+\Omega_{rad})\;. \eea
 Here
$\Omega_{qui}=\rho_{qui}/\rho$ and $\Omega_{rad}=\rho_{rad}/\rho$.

 In the following we will discuss the statefinder for the quintom model with several
kinds of potentials. Firstly we assume that there is no direct
coupling between the phantom scalar field and the normal scalar
field with such a potential $V (\phi, \varphi) = V_{\phi0}
e^{-\alpha \phi} + V_{\varphi0}e^{-\beta\varphi}$, where $\alpha$
and $\beta$ are constants. In this case  the universe has the
phantom dominated late time big rip attractor~\cite{Guo}. In
Fig.~\ref{fig:1} we show the time evolution of statefinder pair
$\{r,s \}$  in  the time interval $\frac{t}{t_0}\in[0.5,4]$ where
$t_0$ is the present time. The model parameters are chosen as
$V_{\phi0}=0.3\rho_0$ and $V_{\varphi0}=0.6\rho_0$, where $\rho_0$
is the present energy density of our universe. We see that in the
past and future the $r-s$ is  almost linear, which means that the
deceleration parameter changes from one constant to another nearly
with the increasing of time, and  the parameters will pass the
fixed point of LCDM in the future.  These  trajectories of  $r(s)$
are different from other dark energy models discussed in
Refs.~\cite{Sahni,Alam, WuYu, Gori, Zimd, Zhang, Zhangx}.
\begin{figure}[htbp] \label{fig:1}
\includegraphics[width=8cm]{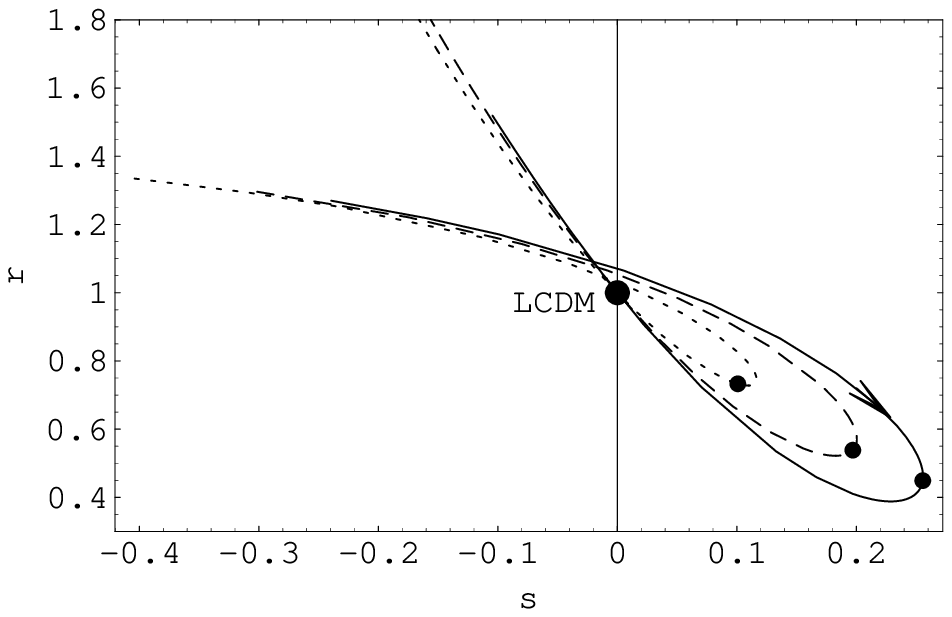}\includegraphics[width=8cm]{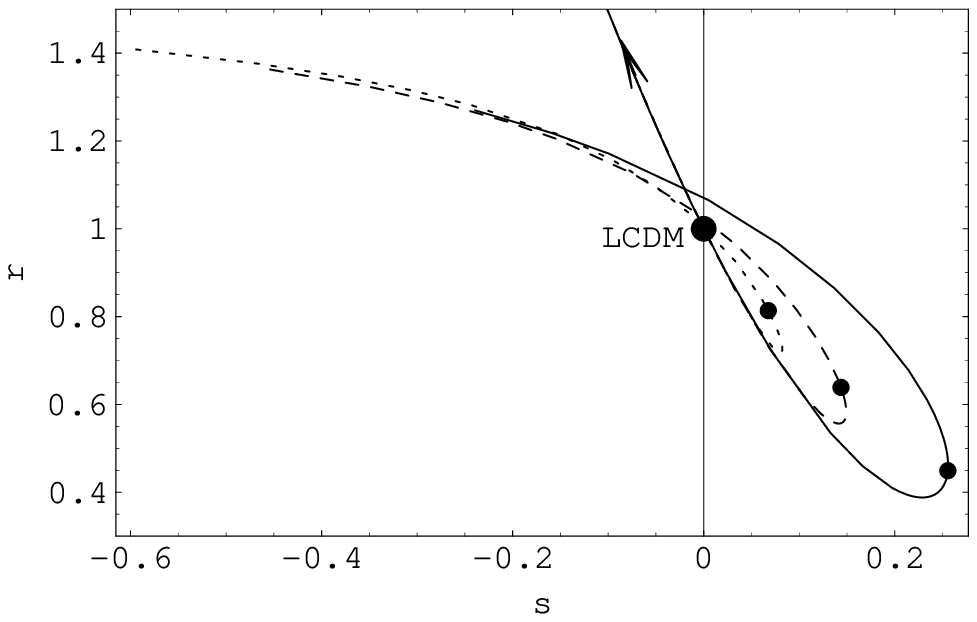}
\caption{\label{fig:1}The $r-s$ diagram of no direct coupling
exponential potential $V (\phi, \varphi) = V_{\phi0} e^{-\alpha
\phi} + V_{\varphi0}e^{-\beta\varphi}$. Curves $r(s)$  evolve  in
the time interval $\frac{t}{t_0}\in[0.5,4]$ where $t_0$ is the
present time. In the left figure
 $\beta=3$ and $\alpha=1, ~1.5,~ 2$(solid line, dashed line and dot-dashed line
 respectively). In the right figure
 $\alpha=1$ and $\beta=2,~ 2.5,~ 3$(solid line, dashed line and dot-dashed line
 respectively). Dots locate the current values of the statefinder parameters. }
\end{figure}

\begin{figure}[htbp]
\includegraphics[width=8cm]{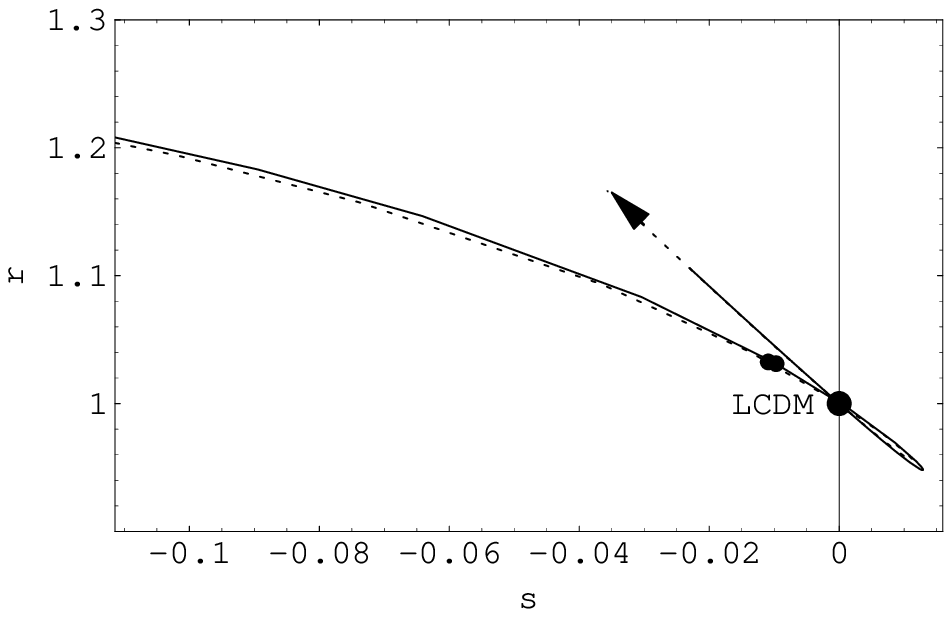}\includegraphics[width=8cm]{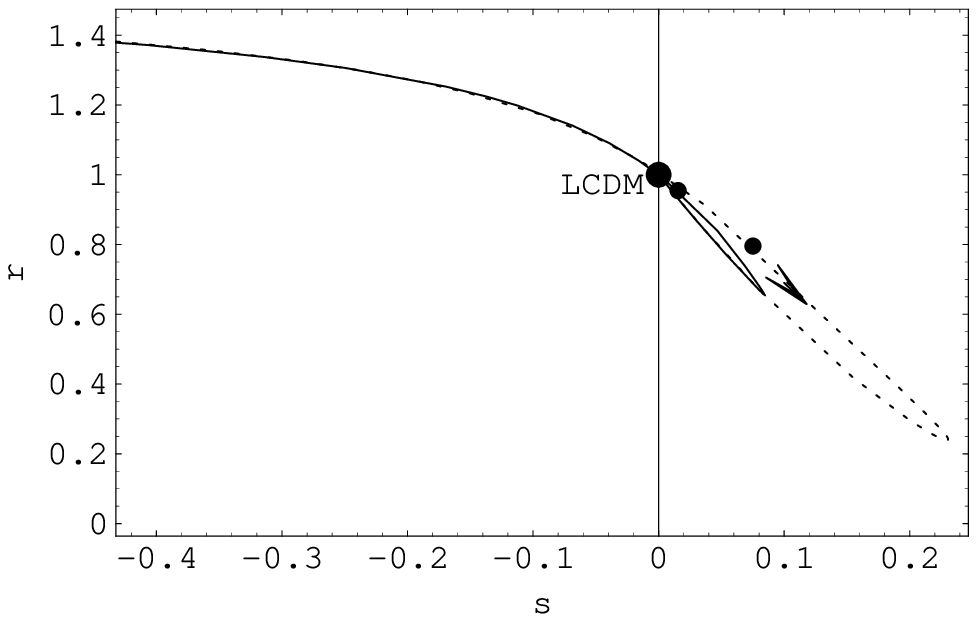}
\caption{\label{fig:2}In the left figure curves $r(s)$  evolve  in
the time interval $\frac{t}{t_0}\in[0.5,4]$ where $t_0$ is the
present time for the potential $V (\phi, \varphi) = V_{\phi0}
e^{-\alpha \phi} +
V_{\varphi0}e^{-\beta\varphi}+V_0e^{-\kappa(\phi+\varphi)}$. The
model parameter are chosen as  $V_{\phi0}=0.3\rho_0$,
$V_{\varphi0}=0.6\rho_0$, $V_{0}=0.3\rho_0$, $\alpha=1$, $\beta=1$
and $\kappa=1,~ 2$(solid line and dot-dashed line
 respectively). In the right figure the potential is  $V
(\phi, \varphi) = V_{\phi0} e^{-\alpha \phi^2} +
V_{\varphi0}e^{-\beta\varphi^2}$. The model parameter are chosen
as $V_{\phi0}=0.3\rho_0$,  $V_{\varphi0}=0.6\rho_0$, $\alpha=1$
and
 $\beta=1,~3$(solid line and dot-dashed line
 respectively). Dots locate the current values of the statefinder parameters. }
\end{figure}

 Next,  two cases of  potential $V (\phi, \varphi) =
V_{\phi0} e^{-\alpha \phi} +
V_{\varphi0}e^{-\beta\varphi}+V_0e^{-\kappa(\phi+\varphi)}$ and $V
(\phi, \varphi) = V_{\phi0} e^{-\alpha \phi^2} +
V_{\varphi0}e^{-\beta\varphi^2}$ will be studied, as before where
$\alpha,~\beta$ and $\kappa$ are constants. The quintom model with
 both above potentials have  the late time attractors but the former
potential leads to a big rip attractor and the latter to a de
Sitter attractor~\cite{ZhangLi, Guo}. In Fig.~\ref{fig:2} the
curves of $r(s)$ are given. Apparently the left figure is very
similar to Fig.~\ref{fig:1}. This shows that for uncoupling and
coupling exponential potentials the evolutions of our universe are
very similar in the time interval we consider here.  The right
figure is different from Fig.~\ref{fig:1} but has a common
characteristic with the phantom with power law
potential~\cite{WuYu}, quintessence with inverse power law
potential and Chaplygin gas model~\cite{Alam} that it reaches the
point of LCDM with the increasing of time.  The reason is that
they all lead to  the same fate of the universe--de Sitter
expansion, but the trajectories to LCDM are different,  therefore
they can be differentiated.

\begin{figure}[htbp]
\includegraphics[width=8cm]{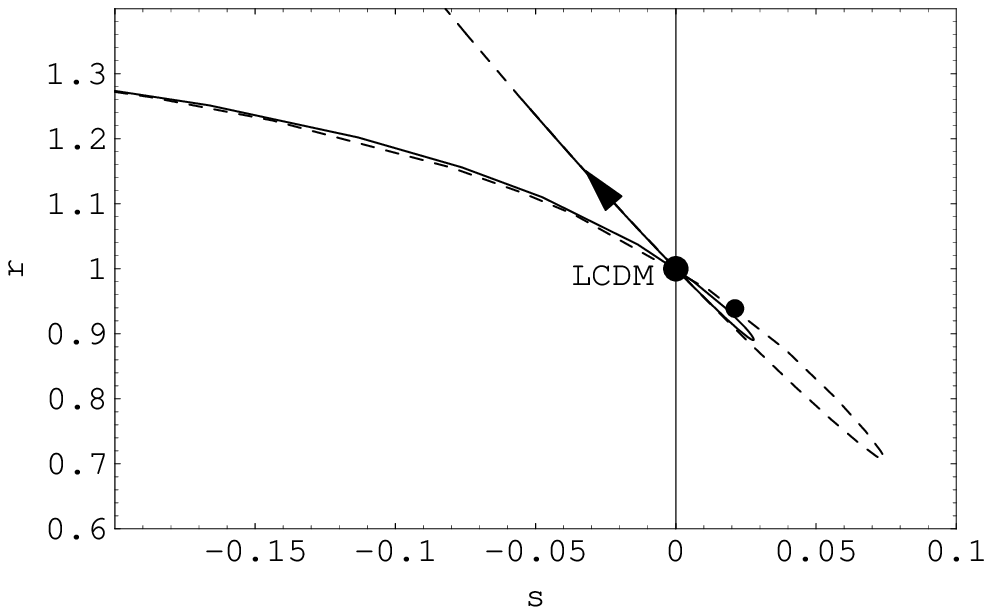}\includegraphics[width=8cm]{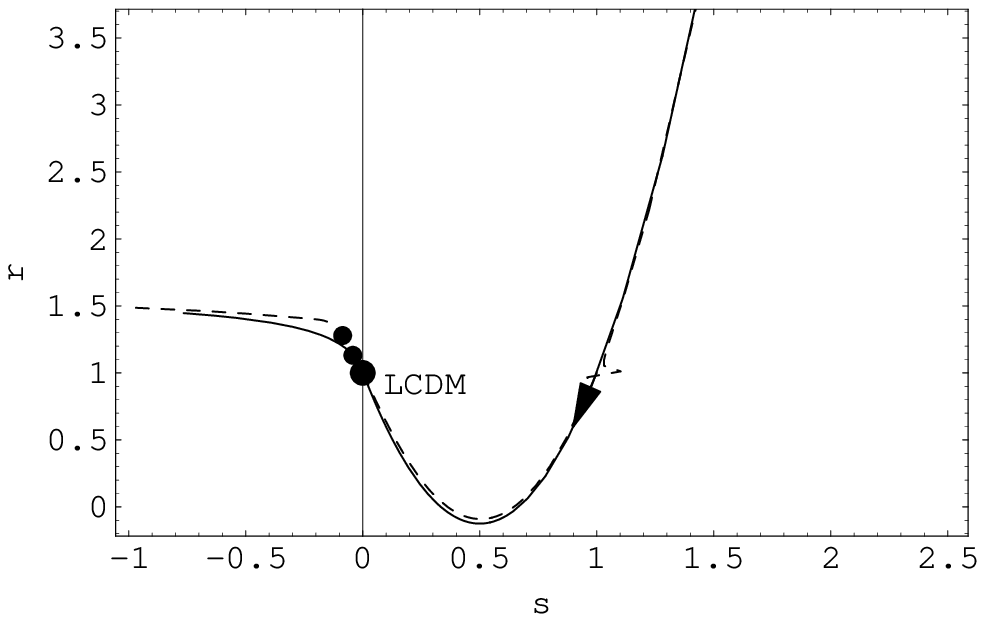}
\caption{\label{fig:3}The $r-s$ diagrams of linear  potential
case. Curves $r(s)$  evolve in the time interval
$\frac{t}{t_0}\in[0.5,4]$. In the left figure
 $\kappa=0.8$ and $\lambda=-0.2,~ -0.8$(solid line and dashed line
 respectively). In the right figure
 $\kappa=0.8$ and $\lambda=0.2,~ 0.8$(solid line and dashed line
 respectively). Dots locate the current values of the statefinder parameters. }
\end{figure}

\begin{figure}[htbp]
\includegraphics[width=8cm]{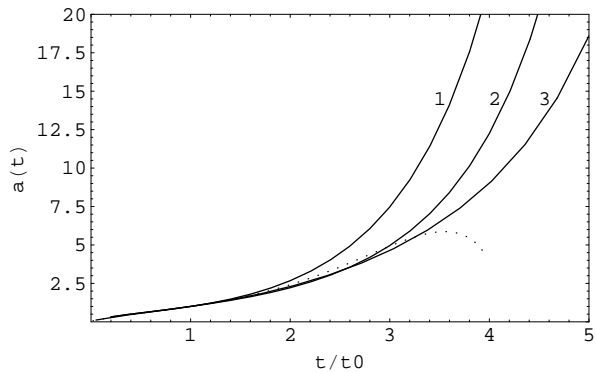}
\caption{\label{fig:4}The evolution of scale factor $a$ with
$t/t_0$. $1$ and $2$ represent potentials $V (\phi, \varphi) =
V_{\phi0} e^{-\alpha \phi} + V_{\varphi0}e^{-\beta\varphi}$ and $V
(\phi, \varphi) = V_{\phi0} e^{-\alpha \phi} +
V_{\varphi0}e^{-\beta\varphi}+V_0e^{-\kappa(\phi+\varphi)}$
respectively. $3$ and dot dashed line denote the linear potential
with negative and positive  coupling respectively. }
\end{figure}

\begin{figure}[htbp]
\includegraphics[width=8cm]{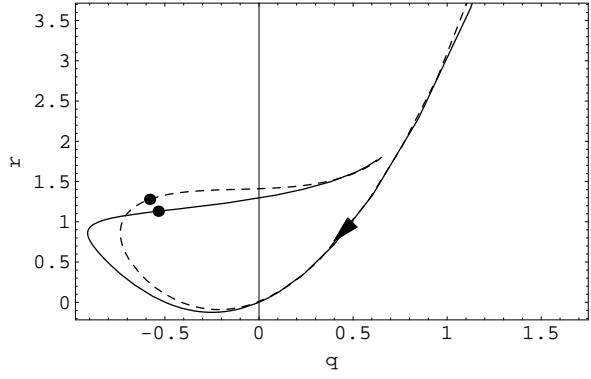}
\caption{\label{fig:5}The $r-q$ diagram of linear  potential case.
Curves $r(q)$  evolve in the time interval
$\frac{t}{t_0}\in[0.2,4]$,
 $\kappa=0.8$ and $\lambda=0.2, 0.8$(solid line and dashed line
 respectively). Dots locate the current values of the statefinder parameters. }
\end{figure}

Finally, we turn to  the  case of linear coupling potential $V
(\phi, \varphi) =\kappa( \phi+\varphi) +\lambda\phi\varphi$, where
$\kappa$ and $\lambda$ are two constants. The scalar field with a
linear potential was firstly studied in Ref.~\cite{Garr} and it
has been argued that such a potential is favored by anthropic
principle considerations~\cite{Garriga,Garriga1,Garriga2} and can
solve the coincidence problem~\cite{Avel}. In addition if the
universe is dominated by quintessence(phantom)  with this
potential it ends with a big crunch(big rip)~\cite{Peri}. The time
evolutions of statefinder pair $\{r,s \}$ in  the time interval
$\frac{t}{t_0}\in[0.5,4]$ are shown in Fig.~\ref{fig:3}. We see
that in the case of negative coupling the diagram is very similar
to Fig.~\ref{fig:1} and the left figure in Fig.~\ref{fig:2}, which
shows that in these cases the evolutions of our universe are
similar in the time interval we consider here,  as shown in
Fig.~\ref{fig:4}. But for $\lambda>0$ the figure of $\{r,s \}$ is
very different. In order to well explain this we give the time
evolutions of $\{r,q\}$ in Fig.~\ref{fig:5}. From Fig.~\ref{fig:4}
and Fig.~\ref{fig:5} we find if the coupling constant $\lambda$ is
positive the accelerating expansion of our universe is temporal
and the universe will decelerate again.

In summary, we study the statefinder parameters of the quintom
dark energy model. Several kinds of potentials are discussed. It
is found that  the statefinder diagnostic can differentiate the
quintom model with other dark energy models, but seems not very
helpful to differentiate quintom dark energy models with some
different kinds of potentials which lead to a similar evolution of
our unverse in the time interval we consider.

\begin{acknowledgments}
 This work was supported in part by the National
Natural Science Foundation of China  under Grants No. 10375023 and
No. 10575035, the Program for NCET under Grant No. 04-0784 and the
Key Project of Chinese Ministry of Education (No. 205110).
\end{acknowledgments}

\end{document}